\documentclass[a4paper,fleqn]{cas-dc}

\usepackage[numbers]{natbib}
\usepackage[version=4]{mhchem}
\usepackage{siunitx}
\usepackage{subfigure}
\def\tsc#1{\csdef{#1}{\textsc{\lowercase{#1}}\xspace}}
\tsc{WGM}
\tsc{QE}
\tsc{EP}
\tsc{PMS}
\tsc{BEC}
\tsc{DE}
\begin{document}
\let\WriteBookmarks\relax
\def\floatpagepagefraction{1}
\def\textpagefraction{.001}
\shorttitle{Phantom LAM and LLI}
\shortauthors{Mohammed Asheruddin N}

\title [mode = title]{Phantom LAM and LLI: Resistance and Hysteresis Bias in Voltage-Curve Degradation Mode Analysis}                      
%\tnotemark[1,2]

%\tnotetext[1]{This document is the results of the research
%   project funded by the National Science Foundation.}

%\tnotetext[2]{The second title footnote which is a longer text matter
%   to fill through the whole text width and overflow into
%   another line in the footnotes area of the first page.}

\author[1]{Mohammed Asheruddin N}[type=author,
                        %auid=000,bioid=1,
                        %prefix=Sir,
                        %role=Researcher,
                        orcid=0000-0002-5082-7562
                        ]
\cormark[1]

%\ead{mnazeeru@ic.ac.uk}
%\ead[url]{www.jkkrishnan.in}

%\credit{Conceptualization of this study, Methodology, Software}

%\address[1]{, Street 129, 1043 NX Amsterdam, The Netherlands}
\affiliation[1]{organization={Imperial College London},
                addressline={Exhibition Rd, South Kensington}, 
                city={London},
               citysep={}, % Uncomment if no comma needed between city and postcode
                postcode={SW7 2AZ}, 
                %state={Kerala},
                country={England}
                }

%\author[2,4]{Han Thane}[style=chinese]

\author[1]{Matheus Leal De Souza}[orcid=0009-0001-0922-6837]
\author[1]{Thomas Holland}[orcid=0009-0009-0670-1901]
\author[1]{Catherine Folkson}[orcid=0009-0000-5121-0874]
\author[1]{Gregory Offer}[orcid=0000-0003-1324-8366]
\author[1]{Monica Marinescu}[orcid=0000-0003-1641-3371]
\cortext[1]{Corresponding author. E-mail address: mnazeeru@ic.ac.uk}

\begin{abstract}
Degradation mode analysis (DMA) is widely used to decompose capacity fade into loss of lithium inventory (LLI) and loss of active material (LAM) from low-rate voltage--capacity data. In practice, however, the measured trace is a pseudo-OCV (pOCV) that retains two non-degradation contributions: (i) an SOC-dependent ohmic drop and (ii) intrinsic charge--discharge hysteresis, particularly in graphite--silicon oxide (C/SiOx) negative electrodes. Here we show that these effects can dominate DMA attribution and generate \emph{phantom LAM/LLI}---apparent material loss created by curve registration, branch choice, and voltage-windowing rather than true degradation. Using two commercial 21700 cells (LG M50T: higher resistance; Molicel P45B: lower resistance), we measure an SOC-dependent instantaneous resistance $R_\Omega(\mathrm{SOC})$ from the first $\sim$50\,ms pulse step and apply an \emph{ohmic-only} IR correction to pOCV prior to fitting. In the LG M50T, IR correction lifts the low-rate discharge pOCV by approximately $+13$--$27$\,mV with ageing and exposes systematic misallocation when IR is ignored: PE-LAM is increasingly under-diagnosed (down to $-8.80\%$ relative error at late life) and LLI is suppressed (median $-3.07\%$), with compensating inflation of apparent graphite loss. In the low-resistance P45B, we isolate hysteresis and window effects on a branch-fair $3.0$--$4.2$\,V window: at end-of-life, the charge branch reports higher PE-LAM ($+3.42$\,pp) and higher LLI ($+5.36$\,pp), while the discharge branch recovers substantially larger Si-LAM (discharge--charge difference rising to $+14.38$\,pp). Truncating the lower cutoff ($2.5$--$4.2 \rightarrow 3.0$--$4.2$\,V) further under-reports Si-LAM by $13.61$\,pp by removing the Si-sensitive low-voltage tail. These results motivate a practical DMA protocol for Gr/Si systems: correct \emph{only} the instantaneous ohmic term, enforce a harmonized voltage window across branches, and base quantitative attribution primarily on the discharge branch, treating anomalous/negative component LAMs on charge as allocation artefacts rather than physical recovery.

\end{abstract}

%\includegraphics{figs/cas-grabs.pdf}

\begin{highlights}
\item ``Phantom'' LAM/LLI arises when DMA is applied directly to uncorrected pseudo-OCV (pOCV) curves.
\item Instantaneous $R_\Omega(\mathrm{SOC})$ from $\sim$50\,ms pulses provides a clean ohmic-only IR correction for DMA.
\item In LG M50T, omitting IR correction suppresses PE-LAM and LLI and inflates apparent graphite loss.
\item In Molicel P45B, hysteresis and voltage-windowing drive large charge--discharge differences in inferred Si loss.
\item A practical prescription: ohmic-only correction, harmonized window, and discharge-branch DMA for robust attribution.
\end{highlights}

\begin{keywords}
Degradation mode analysis (DMA) \sep resistance ($R_\Omega$) correction \sep voltage hysteresis \sep voltage-window sensitivity \sep branch selection \sep graphite--silicon oxide (C/SiOx) anode \sep loss of lithium inventory (LLI) \sep loss of active material (LAM)
\end{keywords}

\maketitle

\section{Introduction}

Lithium-ion batteries now underpin electrified transport and grid storage, placing a premium on diagnostics that can disentangle how capacity and power fade accumulate in operation. Among voltage-based tools, degradation mode analysis (DMA) has proved attractive because it can separate loss of lithium inventory (LLI) from loss of active material (LAM) using only standard cycling data, often aided by differential analyses (dQ/dV or dV/dQ) \cite{1,2,3,4}. Conceptually, this relies on a simple idea: the full-cell signature at a defined temperature should be readable as the difference between positive- and negative-electrode potentials at or near equilibrium, obtained from their respective voltage-capacity traces respective to the Li reference electrode at low current. Therefore, changes in electrodes overlap and shape can be mapped to lithium loss and active-material loss \cite{12,3}.

In practice, the voltage trace available to DMA is not a true equilibrium OCV but a low-current, finite-rest pseudo-OCV (pOCV), ideally under minimal temperature variation. Two distinct departures from equilibrium make this substitution non-trivial. The first is resistance/polarization, a measurement-level translation produced by ohmic drop and mild kinetic/transport losses during galvanostatic operation; it depresses discharge voltage and elevates charge voltage and, if left uncorrected, shifts the entire curve by tens of millivolts relative to the thermodynamic baseline. Early DMA practice mitigated this by operating at very low currents and treating residual polarization as negligible; later work showed that finite polarization still biases peak positions and shapes and argued for explicit compensation or ancillary measurements (e.g., pulse IR, EIS, or pulse-relaxation OCV) before drawing mechanistic conclusions \cite{1,5,6}.

The second is voltage hysteresis, which is intrinsic to the materials’ thermodynamics and persists even after long rests. The pOCV on charge and on discharge differ because the system occupies different metastable configurations and stress states along the two paths. For graphite, a residual $\sim$10–30 mV OCV gap arises from different phase-succession pathways and stacking meta-stability; for Si-containing anodes, hysteresis can be an order of magnitude larger and is strongly coupled to mechanical expansion, plasticity, and slow relaxation \cite{7,8,9}. In blends, this asymmetry propagates to the full-cell pOCV and the derived differential spectra, making branch choice (charge vs. discharge) consequential for DMA. Recent studies have emphasized measuring or modeling hysteresis explicitly-either by constructing quasi-equilibrium pOCV via pulse–relax sequences or by using separate charge and discharge OCV functions (or compact hysteresis models) in analysis-rather than averaging the two branches \cite{10,11}.

These two departures-polarization and hysteresis-interact with how DMA is typically executed. Because DMA is performed within fixed full-cell voltage windows, any uniform translation from resistance changes where the curve intersects the window (re-registering features and altering apparent plateaus), while branch-dependent hysteresis changes the apparent capacity available at a given cut-off and shifts dQ/dV landmarks differently on charge and discharge. The net effect is the risk of apparent LAM and apparent LLI-what we term “phantom” LAM/LLI-arising from curve registration and branch selection rather than genuine deactivation or lithium loss. The literature acknowledges each ingredient separately-near-equilibrium protocols for DMA, sensitivity of differential features to polarization, and the material dependence of hysteresis (especially for graphite staging and Si alloying)-but there is no standardized methodology that (i) corrects the measurement-level resistance translation and (ii) manages the thermodynamic branch dependence coherently within DMA \cite{1,2,3,5,7,8,9,10,11}.

This gap matters more as chemistries diversify, as electrode engineering turns more complex and as power-fade accompanies capacity-fade \cite{13,14,15,16}. Comprehensive reviews on Li-ion degradation stress that diagnostic conclusions depend on test protocol, window, and data processing, yet guidance on resistance correction and hysteresis handling in voltage-curve DMA remains qualitative \cite{2,4,12}. Recent method papers on near-equilibrium curve acquisition and on compact hysteresis models show that both challenges are tractable, but their integration into routine DMA workflows is incomplete \cite{10,11}.

Motivated by this landscape, we ask: how do internal resistance and voltage hysteresis, separately and together, bias voltage-curve DMA (V–Q, dQ/dV, dV/dQ), and what practical protocol-spanning resistance correction, branch selection, and windowing-yields LLI/LAM attributions that best reflect material reality? We address this by formalizing resistance as a reversible measurement translation to be corrected prior to analysis, treating hysteresis as an inherent thermodynamic property to be quantified and managed (not “subtracted”), and examining the sensitivity of commonly used differential metrics to each effect in a way that can be standardized for routine use \cite{1,2,3}.

\section{Methods}

\subsection{Cells, operating protocols, and pseudo-OCV acquisition}

Two commercial 21700 cells were studied to quantify the influence of resistance and hysteresis on voltage-curve degradation mode analysis (DMA): a Molicel P45B (high-Ni NCA positive electrode, graphite-silicon oxide (\ce{C/SiO_x}) negative electrode; power-oriented, low intrinsic resistance) and an LG M50T (NMC811 positive electrode, \ce{C/SiO_x} negative electrode; energy-oriented, higher intrinsic resistance; dataset after Kirkaldy et al.). All tests used a fixed voltage window of 2.5–4.2 V. Unless noted, reference performance tests (RPTs) and diagnostics were performed at 25 \textdegree C in temperature-controlled chambers. Each cell underwent five break-in cycles before life testing to $\leq$ 80 \% capacity retention.

The Molicel cells were aged with 2C constant-current/ constant-voltage (CC–CV) charge, constant-voltage taper to C/20, and 0.5C discharge. The LG M50T cells were aged with 0.3C CC–CV charge, taper to C/100, and 1C discharge. At scheduled RPTs, low-rate galvanostatic traces were recorded as directional pseudo-OCV (pOCV) surrogates: C/20 for Molicel (1 s sampling) and C/10 for M50T (0.1 s sampling). “Charge” denotes lithiation of the negative electrode; “discharge” denotes delithiation. For each RPT, state-of-charge (SoC) was referenced to that RPT’s pOCV discharge capacity (cycle-local SoC). These pOCV curves approximate equilibrium while retaining path dependence (hysteresis), which is subsequently quantified after removal of measurement-level polarization.

\subsection{Resistance characterisation and IR correction}

Polarisation was quantified from embedded pulse sequences in the RPTs. For Molicel, 1C pulses were applied at 25 \%, 50 \% and 75 \% SoC in both directions; for LG M50T, a C/2 GITT discharge comprised 25 pulses transferring 200 mAh each with 1 h rests, down to 2.5 V (the full train executed irrespective of reaching cutoff). The instantaneous ohmic resistance at SoCk  was obtained from the first $\approx$ 50 ms voltage step $\Delta V^*$ after current application, using the signed galvanostatic current ($I>0$ on discharge; $I<0$ on charge):
\begin{equation}
R_\mathrm{\Omega}\ (SoC_k)=\frac{\mathrm{\Delta}V^\ast}{I_{pulse}}
\end{equation}
Subsequent relaxation over 1–60 s was fit with one- or two-exponential terms to report time constants ($\tau_1$, $\tau_2$) as kinetic indicators; these slower contributions were not used for baseline correction to avoid importing rate-dependent structure into the pOCV. Discrete $R_\Omega$  (SoCk) values were interpolated with a shape-preserving cubic Hermite spline to obtain a continuous $R_\Omega$  (SoC). 

\subsection{Hysteresis metrics and voltage-window policy}

Voltage–capacity data from routine performance tests (RPTs) were analysed to quantify cycle-resolved voltage hysteresis between the charge and discharge branches. All analyses used low-rate CC data (constant current segments only; CV/taper samples removed) to minimise rate-dependent curvature and to ensure branch comparability.

Branch-fair comparisons require that charge and discharge be analysed over an identical voltage range. In practice, a complete low-rate discharge (4.20 $\to$ 2.50 V) can be acquired reproducibly after a prior CV charge, whereas a complete charge (2.50 $\to$ 4.20 V) is impractical without an unconventional CV discharge; moreover, finite-current discharge typically relaxes above 2.50 V. For each RPT, we therefore defined a common analysis window $[V_{min}^{com},\ 4.2\ V]$ as the overlap of the charge and discharge voltage domains after trimming the lowest voltages to remove sparse tail samples and outliers. All charge–discharge DMA reported here were computed strictly within this window. In addition, to quantify boundary sensitivity, we performed a discharge-only comparison between the full discharge window (4.20–2.50 V) and a reduced window (e.g., 4.20–3.00 V); the difference in DMA and total hysteresis areas from these two windows serves as a diagnostic for boundary effects.

Raw time-series were mapped to state of charge (SOC) via coulomb counting and normalized by the available discharge capacity in the same RPT (i.e., SOC $\in$ [0,1]over the analysed window). To remove sampling and ordering biases, each branch was (i) sorted monotonically by SOC, (ii) de-spiked and de-naïed (finite-value mask), and (iii) resampled by linear interpolation onto a common SOC grid inside the branch-fair voltage window. The resampling grid was dense ($\geq$ 600 points over the overlap) to ensure numerical stability and was later aggregated into 5 \% SOC windows for reporting.

The local branch separation was defined as
\begin{equation*}
\Delta Vs=V_{chg}(s) - V_{dschg}(s)
\end{equation*}
evaluated on the common SOC grid $s\in[V_{min}^{com},\ V_{max}^{com}]$ implied by $[V_{min}^{com},\ 4.2\ V]$. We quantified hysteresis magnitude as the absolute area between branches, integrated over SOC:
\begin{equation}
A=\Delta Vsds
\end{equation}
reported in $V\cdot\%SOC$ when SOC is expressed in percent. Numerically, A was computed by the trapezoidal rule on the dense grid (which is independent of sampling frequency after resampling). For interpretability, the SOC domain was segmented into 5 \% windows $[s_i,s_{i+1}]$ and window-wise areas
\begin{equation*}
A_i=\int_{s_i}^{si+1} \Delta V(s) | ds
\end{equation*}
are reported alongside window-averaged branch voltages ${\bar{V}}_{chg,i\ }$, ${\bar{V}}_{dchg,i}$. Band totals were formed by summing $A_i$ over a low-SOC band (0–30 \%), a high-SOC band (80–100 \%), and the overall common SOC range.

\subsection{Degradation-mode analysis framework}

We treat the low-rate full-cell curve as a pseudo-OCV (pOCV) and model it as the difference of electrode pseudo-OCPs mapped into the full-cell SoC domain as done by Karger et al. The measured and modeled voltages on the analyzed window are matched by optimizing a small set of geometric (scale/shift) and blend parameters. Full-cell voltage model:
\begin{equation}
U_{meas}(z_{FC})\: \approx\: U_{mod}(z_{FC}; \theta)\: =\: U_{PE} (z_{PE})\: -\: U_{NE} z_{NE}),
\end{equation}
where $U_{PE}$is the positive-electrode pOCP (from half-cell), and $U_{NE}$ is the blend-anode pOCP reconstructed from graphite and silicon references (below). To blend anode (Gr/Si), define electrode-referenced state-of-charge functions (unity at lower cut-off, zero at upper cut-off):
\begin{equation}
{\rm SOC}_{NE}(U)\: =\: \phi_{Si}\: {\rm SOC}_{Si}(U)\: +\: (1-\phi_{Si})\: {\rm SOC}_{Gr}(U),
\end{equation}
with $\phi_{Si}$ the capacity share of Si in the blend. The anode pOCP is the inverse map:
\begin{equation}
U_{NE}(SOC_{NE}) =  [SOC_{NE}(U)]^{-1}.
\end{equation}
Each electrode curve is scaled and shifted to the full-cell coordinate $z_{FC}$ using two parameters per electrode:
\begin{equation}
z_{FC}=\frac{z_{NE}-\sigma_{NE}}{\nu_{NE}},z_{FC}=\frac{(1-z_{PE})-\sigma_{PE}}{\nu_{PE}}.
\end{equation}
The fit parameter vector is
\begin{equation}
\theta=\{\nu_{NE}, \nu_{PE}, \sigma_{NE}, \sigma_{PE}, \phi_{Si}\}.
\end{equation}
On a window with fresh full-cell capacity $Q_{FC}$ (we normalize $Q_{FC}=1$ for convenience),
\begin{equation}
Q_{NE}=\frac{Q_{FC}}{\nu_{NE}},Q_{PE}=\frac{Q_{FC}}{\nu_{PE}},
\end{equation}
\begin{equation}
Q_{LI}=Q_{FC} (\frac{\sigma_{NE}}{\nu_{NE}}+\frac{1-\sigma_{PE}}{\nu_{PE}}).
\end{equation}
Reported modes (fractions on the analyzed window) are
\begin{equation}
{\rm LAM}_x=1-Q_x,x\in{PE, NE},LLI=1-Q_{LI}.
\end{equation}
Component-level anode modes follow from the blend: using the fitted $\phi_{Si}$ and the mappings in (4)–(6), we obtain $Q_{Gr}$ and $Q_{Si}$ over the same window and report
\begin{equation}
{\rm LAM}_{NE,Gr}=1-Q_{Gr},{\rm LAM}_{NE,Si}=1-Q_{Si}.
\end{equation}
The parameters $\theta$ are identified by minimizing a composite error over the window (interior-weighted to de-emphasize end-points), combining voltage residuals and slope residuals:
\begin{equation}
{\min}_\theta\: \parallel U_{meas}-U_{mod}(\theta)\parallel_2\: +\: \lambda\: \parallel\frac{dU_{meas}}{dz_{FC}}-\frac{dU_{mod}(\theta)}{dz_{FC}}\parallel_2.
\end{equation}
As in Karger et al., a small $\lambda$ stabilizes plateau alignment without overweighting noise; details of smoothing and interior masks follow their setup.

\section{Results}

\subsection{Evolution of Resistance and Hysteresis}
\label{sec:3.1}

\begin{figure}
    \centering
    \includegraphics[width=0.7\linewidth]{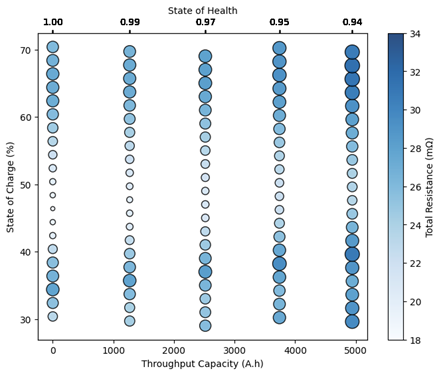}
    \caption{LG M50T resistance evolution during discharge. Instantaneous resistance versus SoC and throughput. Marker color encodes total resistance ($m\Omega$); larger markers at high SoC and late life highlight the SoC-dependent growth. SoH at each RPT is shown above the columns.}
    \label{fig:1}
\end{figure}

\begin{figure*}
    \centering
    \includegraphics[width=0.7\linewidth]{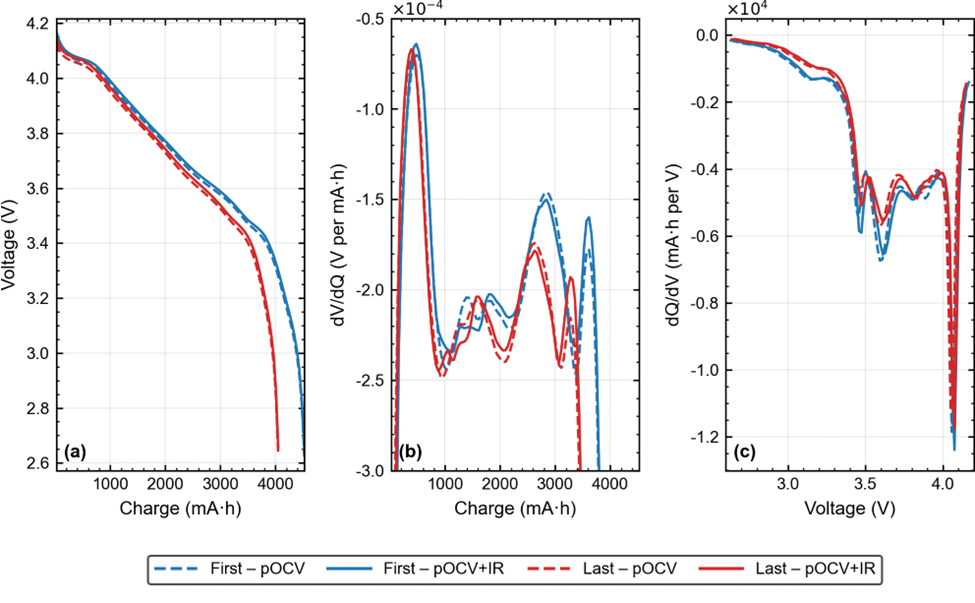}
    \caption{Low-rate signatures before/after IR correction. (a) V–Q: IR correction lifts the curve (blue: first cycle; red: last; dashed = original pOCV, solid = pOCV+IR). (b) dV/dQ–Q: small shape changes; slight spread compaction. (c) dQ/dV–V: consistent right-shifts of peaks/valleys with conserved area (capacity).}
    \label{fig:2}
\end{figure*}

Pulse analysis shows the instantaneous resistance increases with age and toward high SOC (Fig. \ref{fig:1}). Herein, state of health (SOH) was defined as the ratio of discharge capacity in one RPT compared to that of the first one. Across the SoH range 1.00 $\to$ 0.94, $R_\Omega$ rises from roughly 19–22 $m\Omega$  at the earliest RPT to 31–34 $m\Omega$ at the latest, with the largest values above 60 \% SOC. This growth expresses on the low-rate discharge pOCV primarily as an upward translation. After IR correction, as seen in Fig. \ref{fig:2}a, the first-cycle V–Q curve is lifted by a mean +13.2 mV (median +12.0 mV); the lift is +20.1 mV over the upper 10 \% of delivered capacity and +11.4 mV over the lowest 10 \%. By the last cycle, the mean lift is +16.6 mV (median +15.4 mV), with +26.9 mV in the upper 10 \% and +17.0 mV in the lower 10 \%. These gradients indicate that $R_\Omega$  (and any very-fast interfacial component) increases both with aging and near the high-SOC end.

As expected for an algebraic offset, $\mathrm{d}V/\mathrm{d}Q$--$Q$ changes very little (Fig.~\ref{fig:2}b): RMSE between corrected and original is $1.6\times10^{-5}$~V~\si{(mAh)^{-1}} (first cycle) and $2.1\times10^{-5}$~V~\si{(mAh)^{-1}} (last), with median absolute differences around $(5\text{--}6)\times10^{-6}$~V~\si{(mAh)^{-1}}. The spread (IQR) contracts slightly ($-9.1\%$ first, $-6.8\%$ last), consistent with a mild SOC-dependence of the removed polarization and numerical smoothing. In $\mathrm{d}Q/\mathrm{d}V$--$V$ (Fig.~\ref{fig:2}c), capacity is conserved (area unchanged), while spectral features shift to higher voltage: the $\lvert \mathrm{d}Q/\mathrm{d}V\rvert$-centroid moves $+12.9$~mV (first) and $+16.2$~mV (last); the high-$V$ valley ($\sim$4.03--4.18~V) shifts $+18.0$ / $+24.8$~mV, the mid-$V$ valley ($\sim$3.35--3.65~V) $+18.7$ / $+10.4$~mV, and the low-$V$ valley ($\sim$3.05--3.25~V) $+19.3$ / $-4.2$~mV (small, inconsistent change in the last case). These observations confirm that removing the instantaneous IR largely translates the low-rate curve without materially altering its thermodynamic shape.

\begin{figure}
    \centering
    \includegraphics[width=0.7\linewidth]{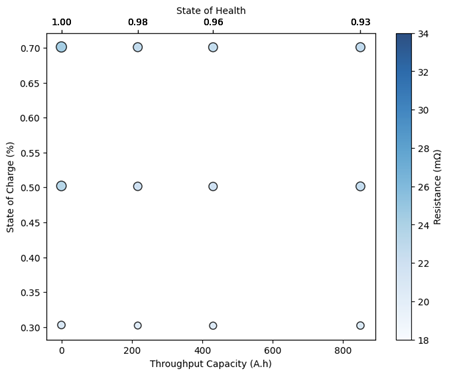}
    \caption{Resistance map (P45B). Instantaneous resistance at $\approx$30, 50, 70 \% SoC across RPTs (SoH $\approx$ 1.00, 0.98, 0.96, 0.93). Points cluster narrowly around 23–26 $m\Omega$  with minimal aging drift, substantially below LG M50T at comparable SoH.}
    \label{fig:3}
\end{figure}

Pulse measurements, Figure \ref{fig:3}, at three SoC levels ($\approx$30, 50 and 70 \%) across four RPTs (SoH $\approx$ 1.00, 0.98, 0.96, 0.93) show that the instantaneous resistance of the Molicel P45B remains low and tightly clustered at 23–26 $m\Omega$  at 25 \textdegree C, with only marginal drift with ageing and weak SoC curvature. Even at the latest RPT the high-SoC points stay within this narrow band and well below the LG M50T values reported in \S \ref{sec:3.1} (19–22 $\to$ 31–34 $m\Omega$  over the same SoH span). This low-IR baseline is consistent with the P45B’s power-cell construction (smaller particles, higher porosity, thinner tortuous domains) and makes the dataset well suited for isolating hysteresis effects without confounding polarization artifacts.

\begin{figure*}
  \centering
  \subfigure[]{
    \includegraphics[width=0.48\linewidth]{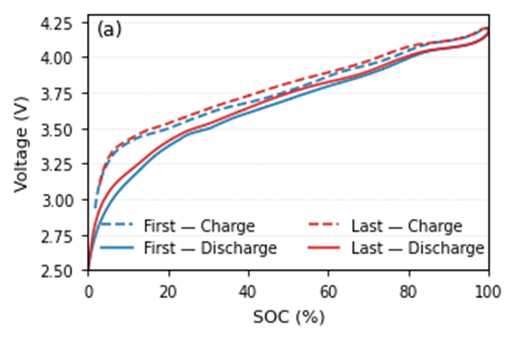}
    \label{fig:4a}
  }
  \hfill
  \subfigure[]{
    \includegraphics[width=0.48\linewidth]{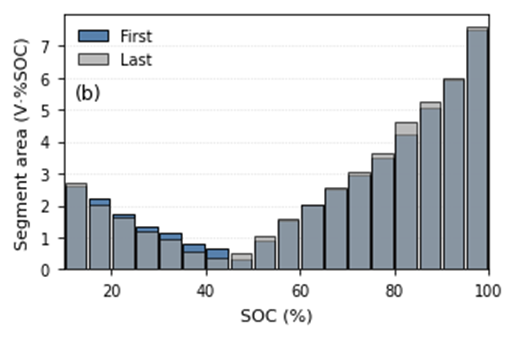}
    \label{fig:4b}
  }
\caption{(a) Low-rate pOCV overlays. Charge (dashed) and discharge (solid) for first and last RPT. The charge trace begins above 2.5 V (non-zero initial SoC) due to relaxation; a common lower-voltage limit is used for fair branch comparison. Separation is smallest near mid-SoC and larger toward both ends. (b) Hysteresis distribution by SoC. Absolute area between branches integrated in 5 \%-SoC windows. Low-SoC contracts, high-SoC grows slightly, and the total over the common window increases marginally.}
\label{fig:4}
\end{figure*}

The low-rate charge/discharge traces in Figure \ref{fig:4a}, exhibit the expected branch separation over the common SoC domain. Because the discharge preceding the charge branch is not followed by an unconventional CV-discharge, the charge trace begins at a voltage above 2.5 V with a correspondingly non-zero initial SoC-the relaxation effect noted in Methods; all branch comparisons therefore use the common overlap. 

Quantifying hysteresis as the absolute area between branches on a common SoC grid and summing in 5 \%-SoC windows; Figure \ref{fig:4b} reveals a clear redistribution with ageing. In the low-SoC region (0–30 \%), hysteresis contracts strongly from 13.406 to 7.374 V·\%SoC ($\Delta$  = -6.032 V·\%SoC, -45 \%). At the high-SoC end (80–100 \%), hysteresis grows slightly from 22.759 to 23.474 V·\%SoC ($\Delta$  = +0.715 V·\%SoC, +3 \%), with the increase concentrated above $\sim$90 \% SoC where the full-cell slope is steep. Integrated over the full common window, the total hysteresis changes little: 47.073 $\to$ 47.427 V·\%SoC ($\Delta$  = +0.353 V·\%SoC, +0.7 \%). The window-wise profile retains a characteristic U-shape at both states of health-small separation near mid-SoC and steeper growth towards the ends-while ageing flattens the left leg ($\approx$0–30 \%) and fattens the extreme right tail ($\gtrsim$ 90 \%).

Taken together, these trends are consistent with a shift in the dominant origin of hysteresis across SoC with ageing: at low SoC the negative-electrode thermodynamic hysteresis associated with Si-rich delithiation (path-dependent phase distributions and stress-coupled equilibria) weakens, producing the pronounced area loss; near the top of charge, modest increases in kinetic/transport penalties (SEI thickening, surface loss, subtle electrolyte/porosity changes) occur precisely where dV/dQ is steep, so small dynamic penalties map into a larger voltage separation and a slightly thicker high-SoC tail. The net result is a nearly conserved total hysteresis with age-dependent reweighting from the low-SoC leg to the high-SoC tail.

For Molicel P45B, resistance remains low (23–26 $m\Omega$) and nearly stationary with ageing, while hysteresis evolves by contracting at low SoC (-45 \%) and thickening slightly at high SoC (+3 \%), leaving the total area essentially unchanged (+0.7 \%). Having established how resistance and hysteresis evolve with ageing in both cells, we now examine how each biases DMA outputs-first the impact of IR correction (\S \ref{sec:3.2}), then the impact of branch choice and voltage window (\S \ref{sec:3.3}).

\subsection{Impact of resistance on DMA}
\label{sec:3.2}
To isolate the diagnostic bias created by polarization, we correct only the ohmic term, which is the sole component that is rate‐proportional and path‐independent across our tests. Dynamic (charge-transfer and concentration) overpotentials depend on history, C-rate and temperature, and the gradients induced by GITT/pulses are not identical to those during a quasi-galvanostatic pOCV trace; attempting to “remove” them would import path dependence. To quantify how much the uncorrected analysis misallocates degradation, we report the relative error of the original (uncorrected) estimate with respect to the IR-corrected one,

\begin{equation}
\varepsilon\ \left[\%\right]=100\left(\frac{Original}{IR-corrected}-1\right)\ 
\end{equation}
so that $\varepsilon\ \left[\%\right]>0$ indicates overestimation by the uncorrected analysis and $\varepsilon\ \left[\%\right]<0$ indicates underestimation. We do not claim the corrected value is the exact thermodynamic truth; it is simply a closer proxy than the raw pOCV because the dominant, rate-proportional drop has been removed. 

\begin{figure*}
    \centering
    \includegraphics[width=1\linewidth]{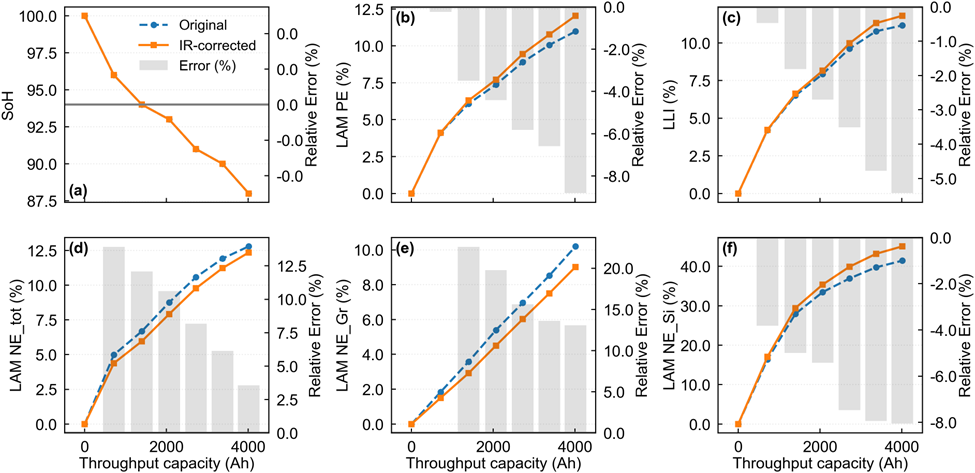}
    \caption{Impact of resistance on DMA-emergence of phantom LAM (LG M50T, 25 \textdegree C, discharge pOCV).Panels show absolute mode values (left axis; Original dashed, IR-corrected solid) and the relative error of the uncorrected estimate with respect to the corrected one (bars, right axis); throughput capacity (Ah) is on the x-axis. (a) SoH; (b) $LAM_{PE}$; (c) $LLI$; (d) $LAM_{NE}$; (e) $LAM_{Gr}$; (f) $LAM_{Si}$}
    \label{fig:5}
\end{figure*}

Applied to the LG M50T energy cell, SoH is, as expected, unchanged by IR correction at every RPT because capacity is taken from low-rate throughput. In contrast, every fitted mode shifts in a systematic, sign-consistent manner that strengthens with ageing, mirroring the known rise of $R_\Omega$ toward high SOC. The cathode active-mass loss is under-diagnosed if IR is ignored: $\varepsilon\ \left[\%\right]LAM_{PE}$ becomes progressively more negative with life, from -0.24 \% at RPT-1 to -8.80 \% at RPT-6 (median $\approx$ -5.12 \%). The negative-electrode total $LAM_{NE}$ shows the opposite sign: $\varepsilon\ \left[\%\right]LAM_{NE}$ is positive - the uncorrected analysis overestimates $LAM_{NE}$ - decreasing from+13.77 \% to +3.64 \% across RPT-1$\to$6 (median $\approx$ +9.38 \%). Decomposing the blend reveals the origin: the graphite component is consistently inflated without IR (median $\approx$ +17.68 \%). In contrast, the silicon component is suppressed without IR (median $\approx$ -6.44 \%). Finally, LLI is likewise under-diagnosedif IR is ignored (median $\approx$ **-3.07 \%**).

These signs and magnitudes follow directly from how a SOC-dependent vertical translation re-allocates the limited “voltage budget” in the mapping. In LG M50T, $R_\Omega$ is largest at high SOC and grows with age; adding ${IR}_\Omega$ lifts the discharge pOCV most strongly at the top end. The cathode pOCP is steep there, so the lift exposes a larger curvature deficit that the optimizer must absorb as greater PE scaling  (larger effective $\nu_{PE}^{-1}$ $\to$ more $LAM_{PE}$); hence the uncorrected fit underestimates PE loss, and the magnitude of that underestimation becomes more negative with life (-0.24 \% $\to$ -8.80 \%). Graphite, by contrast, is dominated by long staging plateaus; a (nearly) uniform downward sag of the full-cell curve from ${IR}_\Omega$ mimics capacity loss on the NE side because the plateau voltage appears “too low” at a given SOC. The optimizer compensates by over-scaling the NE (smaller $\nu_{NE}$) which DMA reads as graphite LAM. When the curve is lifted by IR correction, that mis-scaling is no longer needed and the phantom graphite LAM recedes. Silicon sits on the steep NE segment near the top of discharge; IR sag partially hides that steepness, so the uncorrected fit requires less Si scaling, whereas the corrected fit restores the steepness and reallocates loss to Si-giving a consistently negative $\varepsilon\ \left[\%\right]$ for $LAM_{Si}$. Within the DMA framework, LLI is set by the overlap of the shifted electrode pOCPs; depressing the discharge trace artificially increases this overlap at the voltage-window limits, so removing ${IR}_\Omega$ reduces the overlap and LLI rises, exactly as seen in the monotonic $\varepsilon\ \left[\%\right]$ sequence (-0.37 \% $\to$ -5.43 \%).

In sum, the SOC- and age-dependent ohmic landscape in this high-Ni energy cell generates a characteristic diagnostic artifact when left uncorrected: a large, directional misallocation of degradation that manifests as phantom $LAM_{Gr}$, with concomitant under-diagnosis of PE and Si losses and suppression of LLI. Although IR-corrected DMA remains a proxy (residual kinetic polarization at C/10–C/20 persists), it is unequivocally closer to the thermodynamic picture than raw pOCV and suppresses the systematic, SOC-dependent phantom LAM that otherwise arises from ohmic depression of the discharge curve.

\subsection{Impact of hysteresis on DMA}
\label{sec:3.3}

Voltage hysteresis is an intrinsic, thermodynamic property of the negative electrode (dominantly the Si fraction in these blends), whereas the positive electrode and graphite exhibit comparatively minor hysteresis. Because DMA maps a single full-cell voltage trace onto electrode pOCP templates, any systematic offset between charge and discharge branches can be misread as changes in active mass (LAM) or lithium inventory (LLI). Two experimental choices determine how strongly this bias appears: (i) the voltage window used for the fit, which governs how much of the low-voltage anode “tail” and high-voltage cathode curvature are sampled; and (ii) the branch (charge vs discharge), which differ in their proximity to equilibrium because of path-dependent stress and phase distributions in the Si-containing anode. We therefore (a) harmonize the voltage window to isolate boundary effects, and (b) on that common window, quantify the intrinsic hysteresis bias of charge vs discharge for DMA.

\subsection{Effect of voltage window}

Molicel P45B-by design a low-resistance, high-power 21700 with small secondary particles, high porosity - is a clean platform to isolate voltage-window bias from ohmic artifacts. To quantify that bias, we performed the identical DMA on discharge pOCV traces over two windows that differ only in the lower cutoff: the full discharge window, 2.5–4.2 V, and the common window, 3.0–4.2 V, which is used later for fair charge–discharge comparisons. All fitting settings and electrode pseudo-OCP templates are unchanged; only the sampled voltage span differs. The consequence is systematic and grows with aging: truncating the window consistently under-reports degradation and redistributes negative-electrode loss from Si to graphite because the omitted 2.5–3.0 V band is precisely where the Si-dominated degradation signature resides (deep delithiation under stress, isolation and pore/SEI growth).

\begin{figure*}
    \centering
    \includegraphics[width=1\linewidth]{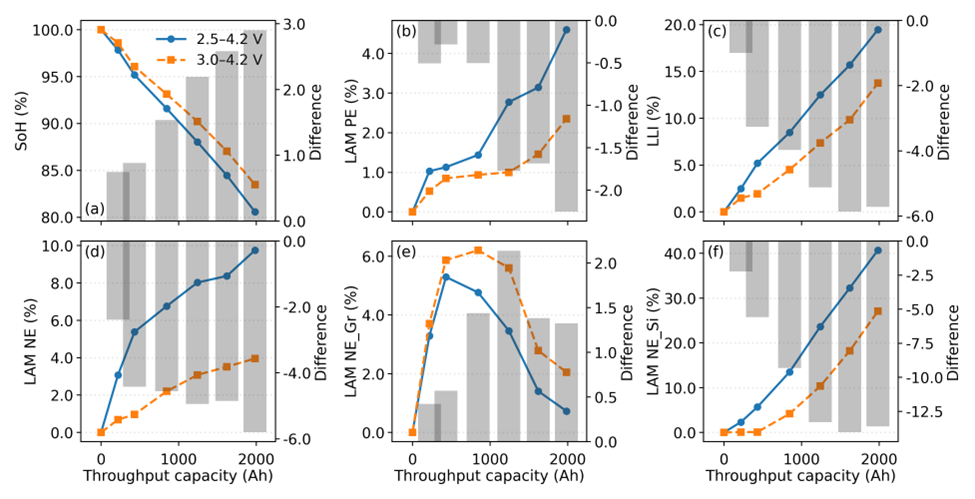}
    \caption{Voltage-window bias in DMA (Molicel P45B, 25 \textdegree C, discharge pOCV).
    Six panels compare 2.5–4.2 V (solid) and 3.0–4.2 V (dashed): (a) SoH, (b) $Q_{PE}$, (c) $Q_{LI}$, (d) $Q_{NE}$, (e) $Q_{Gr}$, (f) $Q_{Si}$. Bars (right axes) show absolute differences in percentage points, (3.0–4.2) - (2.5–4.2).}
    \label{fig:6}
\end{figure*}

Within-window capacity (SoH) is the first signal of this protocol effect. By the last reference performance test (RPT; Throughput = 1 982.5 Ah), SoH increases from 80.57 \% in the 2.5–4.2 V analysis to 83.47 \% in 3.0–4.2 V, a +2.90 percentage-point (pp) lift ($\approx$ +3.6 \% relative); at RPT-1 the lift is already +0.75 pp, and it grows monotonically thereafter (Fig. \ref{fig:6}a). The same “healthier-looking” picture appears in the total negative-electrode retention: $Q_{NE}$ rises from 90.25 \% to 96.04 \% at EOL (+5.80 pp, Fig. \ref{fig:6}d), which in DMA terms means ${LAM}_{NE}$ falls by 5.8 pp purely because the most degraded low-voltage segment is no longer sampled (RPT-1: +2.39 pp). This is not material recovery; it is the direct consequence of what part of the thermodynamic curve is being fitted.

The partition inside the anode explains where that apparent gain comes from. Silicon’s retained fraction, $Q_{Si}$, jumps from 59.29 \% (2.5–4.2 V) to 72.91 \% (3.0–4.2 V) at EOL-+13.61 pp-so ${LAM}_{NE,Si}$ is under-reported by 13.61 pp when the low-voltage tail is omitted (Fig. \ref{fig:6}f). The difference is already visible at RPT-1 (+2.23 pp) and grows cycle-by-cycle, quantitatively showing that the 2.5–3.0 V band carries an increasing share of the Si-specific loss signature with age. Graphite moves in the opposite (small) direction: $Q_{Gr}$ is lower in the truncated window-97.95 \% vs 99.27 \% at EOL (-1.32 pp, Fig. \ref{fig:6}e)-so ${LAM}_{NE,Gr}$ rises by 1.32 pp ($\approx 1.8\times$ the full-window graphite loss). Mechanistically, once the Si lever arm below 3.0 V is removed, the remaining mid-SOC curvature must be matched with a bit more graphite scaling; the fit duly shifts a small portion of NE loss from Si to Gr. Crucially, the graphite shift stays low-amplitude ($\approx$ 1–2 pp), while the silicon shift is large and age-dependent, consistent with Si being the true driver of NE degradation in this blend.

Windowing also relaxes cathode and inventory penalties through the geometry of pOCP overlap at the window bounds. At EOL, $Q_{PE}$ increases from 95.40 \% to 97.65 \% (+2.25 pp, Fig. \ref{fig:6}b), so ${LAM}_{PE}$ is under-reported by 2.25 pp ($\approx$ -49 \% relative to the full-window PE loss). Likewise, $Q_{LI}$ increases from 80.53 \% to 86.25 \% (+5.72 pp, Fig. \ref{fig:6}c), so LLI is under-reported by 5.72 pp ($\approx$ -29 \%). Early-life shifts are smaller but have the same sign (RPT-1: $Q_{PE}$+0.51 pp, $Q_{LI}$+1.00 pp), and both trends grow steadily with throughput. Intuitively, by removing the low-V anode tail you reduce the anode’s constraining leverage and increase apparent pOCP overlap within the truncated span; the optimizer then needs less cathode scaling and assigns less inventory loss to match the same measured voltage.

Altogether, the data show that boundary selection is a first-order source of diagnostic bias for DMA on Gr/Si systems. Moving from 2.5–4.2 V to 3.0–4.2 V makes the cell look uniformly healthier, with the largest under-reporting on Si LAM (-13.6 pp at EOL), a modest over-reporting of graphite LAM (+1.3 pp), and non-negligible under-reporting of PE-LAM (-2.3 pp) and LLI (-5.7 pp), while boosting apparent SoH by +2.9 pp. Because these shifts arise solely from the voltage span, not from kinetics or noise, all subsequent charge–discharge hysteresis comparisons (\S  3.3.2) are performed on the harmonized 3.0–4.2 V window to ensure that any remaining differences are attributable to hysteresis itself, not to boundary selection.

\subsection{Effect of hysteresis}

\begin{figure*}
    \centering
    \includegraphics[width=1\linewidth]{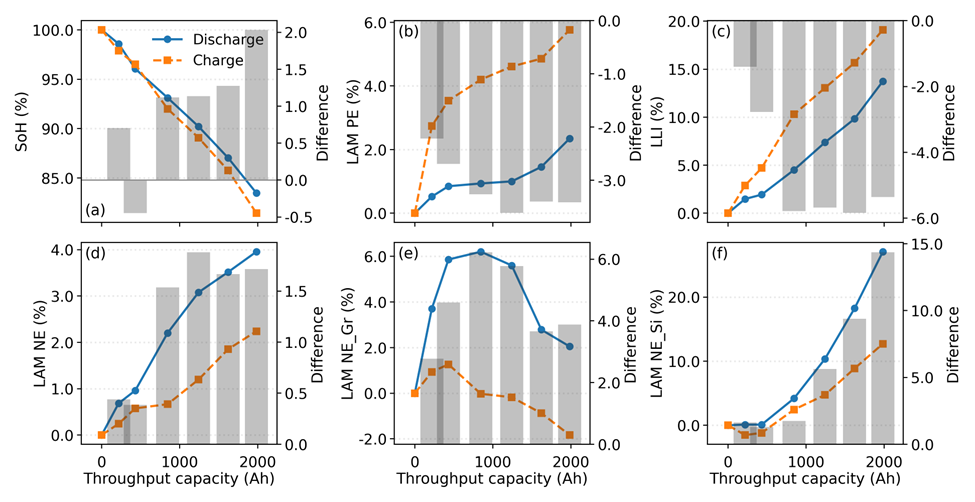}
    \caption{ Charge–discharge hysteresis bias in DMA on a common 3.0–4.2 V window (Molicel P45B, 25 \textdegree C). Blue solid lines: discharge; orange dashed lines: charge. Grey bars (right axes) show Discharge - Charge in percentage points for each metric. Panels: (a) SoH, (b) ${\rm LAM}_{PE}$, (c) $LLI$, (d) ${\rm LAM}_{NE}$ (total), (e) ${\rm LAM}_{NE,Gr}$, (f) ${\rm LAM}_{NE,Si}$. }
    \label{fig:7}
\end{figure*}

To isolate hysteresis from boundary effects, degradation-mode analysis was performed on both branches within the same 3.0–4.2 V pOCV window. The resulting contrasts in Fig. \ref{fig:7} therefore reflect branch-dependent thermodynamics of the Gr/Si negative electrode (NE), with only minor residual kinetic contributions at the low rates used. Two signatures dominate the data. First, at fixed passed charge, the charge branch sits at a higher full-cell voltage than the discharge branch across most of the window. Second, the cathode top-end is steep, so small vertical differences at a given capacity compel appreciable changes in the fitted electrode scalings and the inferred Li-inventory overlap.

Quantitatively, the charge branch persistently reports more cathode loss and more Li-inventory loss, while the discharge branch recovers larger anode loss-especially from silicon. At end-of-life (1982.5 Ah), ${\rm LAM}_{PE}$ is 3.42 pp higher on charge and LLI is 5.36 pp higher on charge; the same direction holds at every RPT and both deltas grow with ageing (Fig. \ref{fig:7}b–c). Conversely, the discharge branch assigns substantially more NE loss to Si: the difference Dis-Chg in ${\rm LAM}_{NE,Si}$ rises almost monotonically from +1.63 pp at the first RPT to +14.38 pp at EOL (Fig. \ref{fig:7}f). A smaller but systematic allocation also appears in graphite: discharge exceeds charge by +2.8–6.2 pp mid-life and +3.9 pp at EOL (Fig. \ref{fig:7}e). Aggregating Si and Gr gives a consistent picture: ${\rm LAM}_{NE}$ is higher on discharge by $\sim$0.4–1.9 pp from mid-life onward (Fig. \ref{fig:7}d). The within-window capacity (SoH) is correspondingly slightly higher on discharge-by +0.7–2.0 pp, save for one early RPT with -0.45 pp (Fig. \ref{fig:7}a)-because the elevated charge voltage reaches the upper cutoff with less area under the curve.

These shifts are the expected consequence of NE hysteresis in a Gr/Si blend and the geometry of the mapping $U_{\mathrm{cell}}=U_{\mathrm{PE}}-U_{\mathrm{NE}}$. During charge, the NE delithiates along a higher-potential path owing to silicon’s stress-coupled, path-dependent quasi-OCV. The measured full-cell curve is therefore uplifted relative to discharge at the same Ah. Fitting this uplift within a fixed window forces (i) greater cathode scaling to reproduce the steep PE curvature near 4.2 V, which appears as higher ${\rm LAM}_{PE}$ on charge; (ii) smaller modeled Li-overlap (via $\sigma$-type horizontal shifts), which appears as higher LLI on charge; and (iii) a re-weighting inside the NE to hold the mid-SOC shape under the elevated voltage, which down-weights Si on charge and shifts a small residual onto graphite. This re-weighting intensifies with ageing: as the true Si lever arm shortens (seen cleanly on the discharge branch), the model’s incentive to depress Si scaling on charge grows, hence the rising $\Delta$ Si-LAM with throughput. Graphite’s role is transient because its long plateaus can mimic curvature early; as Si degradation strengthens, the charge-side need for extra graphite loss relaxes, and ${\rm LAM}_{NE,Gr}$ on charge can even dip slightly negative late-life.

A note on “negative” component LAMs observed on the charge branch at early RPTs (e.g., slightly negative Si-LAM): these are not physical capacity gains. In the blend reconstruction, the NE pOCP is built as a mixture of graphite- and silicon-like responses and then scaled/shifted to match the full-cell curve over a restricted span. When one branch is hysteretic and elevated, the optimizer can achieve the required curvature in 3.0–4.2 V by assigning an effective Si retention in that span slightly above the fresh normalization, i.e., $Q_{Si}^{\mathrm{charge}}>100\%$. Reported as ${\rm LAM}_{NE,Si}=100(1-Q_{Si})$, this appears negative. Two consistency checks hold simultaneously: the total NE retention remains plausible and the blend rule $Q_{NE}\approx(1-\phi_{Si})Q_{Gr}+\phi_{Si}Q_{Si}$ stays within numerical tolerance. The “negative” values should therefore be interpreted as under/over-compensation inside the NE decomposition driven by hysteresis, not as material rejuvenation.

In short, even on a common 3.0–4.2 V window, hysteresis alone biases DMA: charge inflates ${LAM}_{PE}$ and LLI while suppressing Si-LAM, with a transient graphite redistribution; discharge is closer to equilibrium and consistently recovers the Si-dominated NE loss with smaller cathode and inventory penalties. This pattern is precisely the mechanistic footprint of phantom LAM/LLI introduced by analyzing a hysteretic branch, and it motivates our recommendation to base quantitative DMA on the discharge branch whenever possible, reserving charge-branch fits for qualitative cross-checks or for cases where hysteresis itself is the subject of study.

\section{Discussion}

In a low-rate pOCV analysis the measured full-cell curve on a given branch can be written as

\begin{equation*}
U_{\mathrm{meas}}(Q)=U_{\mathrm{PE}}^{(b)}(z_{\mathrm{PE}})-U_{\mathrm{NE}}^{(b)}(z_{\mathrm{NE}})+\eta_{\mathrm{kin}}(z,I)+I R_\Omega
\end{equation*}
with branch-specific pseudo-OCPs $U^{(b)}$ and an instantaneous ohmic term $IR_\Omega$. The DMA then determines the coordinate transforms ($\nu_{NE}$, $\nu_{PE}$, $\sigma_{NE}$, $\sigma_{PE}$) (and the blend fraction $\phi_{Si}$ for Gr/Si) that make $U_{\mathrm{PE}}^{(b)}-U_{\mathrm{NE}}^{(b)}$ reproduce $U_{\mathrm{meas}}$ in the chosen window, mapping directly to $Q_{PE}$, $Q_{NE}$, $Q_{LI}$ and hence LAM and LLI \cite{1,2}. Ohmic resistance behaves to first order as a vertical translation of $U_{\mathrm{meas}}$ (negative on discharge, positive on charge). Because the cathode top-end carries a large $\partial V/\partial Q$, even a modest $\mid I\mid R_\Omega$ (we observe +13–20 mV early, +16–27 mV late after correction) disproportionately alters the fitted $\nu_{PE}$ and pOCP overlap, biasing ${\rm LAM}_{PE}$ and LLI if left uncorrected; the tiny residual in dV/dQ after IR removal confirms that dynamic polarization is small at C/10–C/20 [Kirkaldy et al., 2024] \cite{2}. Accordingly, correcting only $R_\Omega$ (SOC) measured independently-rather than folding history-dependent diffusion overpotentials into the thermodynamic fit-preserves identifiability and aligns with DMA best practice \cite{2,1} .

Hysteresis is intrinsically thermodynamic and branch-dependent for Gr/Si electrodes: during full-cell charge the NE delithiates along a higher-potential path owing to stress-coupled phase transformations in silicon, while discharge follows a lower-potential lithiation path [Kim et al., 2021; Berg et al., 2025]. When both branches are analysed on the same 3.0–4.2 V window, the differences are therefore hysteresis bias rather than boundary effects. Consistent with this, we find that charge systematically inflates ${\rm LAM}_{PE}$ by $\approx$2–3.5 pp and LLI by $\approx$1.4–5.8 pp across RPTs, whereas discharge recovers substantially more NE loss, dominated by silicon (Dis-Chg in ${\rm LAM}_{NE,Si}$ increases from +1.6 pp to +14.4 pp with ageing); a smaller transient reallocation appears in graphite (+2.8–6.2 pp mid-life). Mechanistically, lifting the charge curve forces greater cathode scaling and smaller pOCP overlap to reproduce the steep top-end, hence higher ${\rm LAM}_{PE}$ and LLI; to preserve mid-SOC curvature under that uplift, the solver down-weights the Si-like segment of the NE blend within the restricted span, suppressing Si-LAM on charge-occasionally to slightly negative algebraic values early life. These “negative” component LAMs are not material gains; they reflect under/over-compensation within the NE blend over a narrow window and remain consistent with blend-rule checks $Q_{NE} \approx (1-\phi_{Si})Q_{Gr}+\phi_{Si}Q_{Si}$[Karger et al., 2023]. 

\begin{figure*}
    \centering
    \includegraphics[width=1\linewidth]{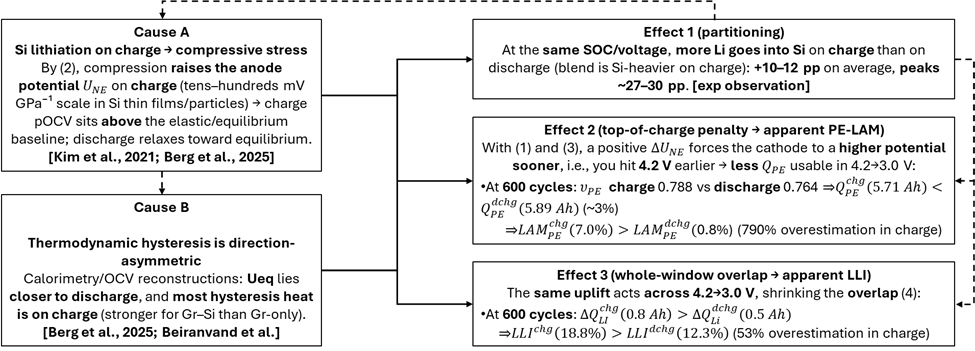}
    \caption{Mechanistic insights into the impact of resistance and hysteresis on DMA}
    \label{fig:8}
\end{figure*}

\begin{figure}
    \centering
    \includegraphics[width=1\linewidth]{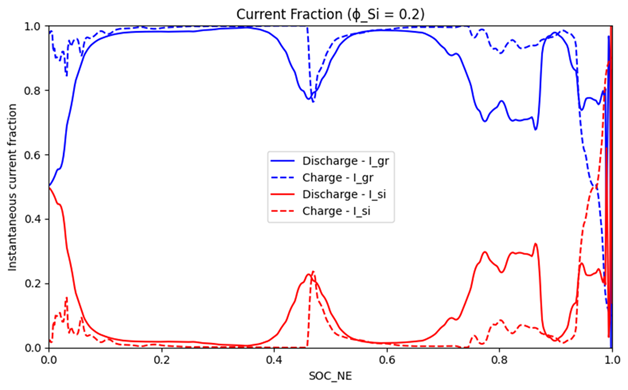}
    \caption{Current partition between Gr and Si in the NE during charge and discharge}
    \label{fig:9}
\end{figure}

Figure \ref{fig:8} usefully summarizes the causal chain: (i) ohmic translation at high-SOC $\to$ parameter changes concentrated in $\nu_{PE}$, $\sigma\to$ phantom under-estimation of ${\rm LAM}_{PE}/LLI$ on discharge unless IR-corrected; (ii) branch uplift from NE hysteresis on charge $\to$ larger $\nu_{PE}$ and smaller pOCP overlap $\to$ phantom over-estimation of ${\rm LAM}_{PE}/LLI$ and under-estimation of Si-LAM on charge. The current-partition plot (Fig. \ref{fig:9}), derived from the blend templates and local slopes, shows that across most of mid-SOC ($\approx$10–70\% NE-SOC) the negative electrode is graphite-dominated (graphite current fraction $>$90\%), with only a small Si “blip” around the graphite staging crossover. As SOC approaches the extremes-where Si becomes thermodynamically active-the branches diverge: on discharge (NE lithiation) the Si share rises smoothly and over a broader SOC band (typically reaching $\sim$10–30\% across several lobes), indicating a more compliant Si response; on charge (NE delithiation) Si participation is suppressed and abrupt, engaging only in narrow windows and dropping back quickly, while graphite continues to carry the bulk of current. This branch asymmetry reflects a higher Si stiffness (larger $\partial U/\partial z$) and elevated potential on delithiation due to stress/phase path-dependence, versus a more relaxed, closer-to-equilibrium Si path on lithiation. Practically, the smoother, more distributed Si participation on discharge preserves the Si-driven curvature of the full-cell pOCV and leads DMA to assign larger Si-LAM with smaller PE-LAM/LLI; the abrupt, reduced Si engagement on charge shifts the fit toward higher PE-LAM and LLI and lower (sometimes negative) Si-LAM near the top of charge.

Two practice-ready points follow: 
\begin{itemize}
    \item Use IR-corrected pOCV and perform quantitative DMA on the discharge branch in a common voltage window; reserve charge-branch DMA for qualitative cross-checks or when hysteresis itself is the target metric.
    \item When reporting charge-branch component LAMs, flag any negative values as allocation artefacts and accompany them with blend-consistency residuals and, if space allows, a constrained re-fit sensitivity \cite{2} [Karger et al., 2023].
\end{itemize}

\section{Conclusion}

Degradation-mode analysis (DMA) only reports the physics that the voltage trace allows it to see. By separating ohmic resistance from branch hysteresis on two complementary 21700 systems and enforcing common-window fits, we showed how each creates phantom LAM/LLI if unaccounted for. First, ohmic drop acts as a near-rigid vertical translator of the pOCV; because the PE top-end is steep, small IR offsets propagate into underestimated PE-LAM and LLI unless corrected. Second, the Gr/Si negative-electrode is intrinsically hysteretic: the charge branch (NE delithiation) sits at higher potential than discharge (NE lithiation), so same-window fitting inflates PE-LAM and LLI on charge while suppressing Si-LAM (sometimes algebraically negative) and modestly reassigning loss from Si to graphite. A current-partition analysis clarifies the origin: Si carries a smoother, broader share on discharge but is stiffer and engages abruptly on charge, exactly where PE slope maximizes parameter leverage. Window choice further modulates attribution by omitting (or including) the Si-sensitive low-V tail.

Practical prescription: correct only the ohmic term prior to fitting, perform quantitative DMA on the discharge branch within a harmonized voltage window, and treat charge-branch component LAMs as allocation artefacts when negative or anomalously small.

Crux: If the aim is to quantify materials loss rather than protocol artefact, the most faithful configuration is IR-corrected, discharge-branch DMA on a common window-it consistently recovers Si-dominated NE loss without over-penalizing the cathode or the lithium inventory.

\section*{CRediT authorship contribution statement}
Mohammed Asheruddi N: Conceptualization, Methodology, Validation, Formal analysis, Investigation, Data curation, Writing – original draft, Writing – review \& editing, Visualization. Matheus Lead De Souza: Validation, Writing – review \& editing, Visualization. Catherine Folkson: Investigation, Data curation. Gregory Offer: Conceptualization, Methodology, Validation, Formal analysis, Resources, Writing – review \& editing, Visualization, Supervision, Project administration, Funding acquisition. Monica Marinescu: Conceptualization, Methodology, Validation, Formal analysis, Resources, Data curation, Writing – review \& editing, Visualization, Supervision, Project administration, Funding acquisition.

\section*{Declaration of competing interest}
The authors declare that they have no known competing financial
interests or personal relationships that could have appeared to influence
the work reported in this paper.

\section*{Acknowledgements}
The authors gratefully acknowledge financial support from Innovate UK (MESM\_PA8435) and The Faraday Institution’s Multiscale Modelling project (MESM\_PB3785).

\section*{Data availability}
The data supporting the findings of this study are publicly available on Zenodo
(\url{https://zenodo.org/records/10637534}).
The degradation mode analysis (DMA) tools used in this work are available as open-source
software in the PyProBE repository
(\url{https://github.com/ImperialCollegeLondon/PyProBE}).

\bibliographystyle{cas-model2-names}
\bibliography{cas-refs}

@article{1,
  author  = {Dubarry, Matthieu and Truchot, Christophe and Liaw, Bor Yann},
  title   = {{Synthesize battery degradation modes via a diagnostic and prognostic model}},
  journal = {Journal of Power Sources},
  volume  = {219},
  pages   = {204--216},
  year    = {2012},
  doi     = {10.1016/j.jpowsour.2012.07.016}
}

@article{2,
  author  = {Birkl, Christoph R. and Roberts, Matthew R. and McTurk, Edward and Bruce, Peter G. and Howey, David A.},
  title   = {{Degradation diagnostics for lithium ion cells}},
  journal = {Journal of Power Sources},
  volume  = {341},
  pages   = {373--386},
  year    = {2017},
  doi     = {10.1016/j.jpowsour.2016.12.011}
}

@article{3,
  author  = {Dubarry, Matthieu and Anse{\'a}n, David},
  title   = {{Best practices for incremental capacity analysis}},
  journal = {Frontiers in Energy Research},
  volume  = {10},
  pages   = {1023555},
  year    = {2022},
  doi     = {10.3389/fenrg.2022.1023555}
}

@article{4,
  author  = {Olson, Jonathan Z. and Lopez, Carlos M. and Dickinson, Edward J. F.},
  title   = {{Differential analysis of galvanostatic cycle data from Li-ion batteries: Interpretative insights and graphical heuristics}},
  journal = {Chemistry of Materials},
  volume  = {35},
  number  = {4},
  pages   = {1487--1513},
  year    = {2023},
  doi     = {10.1021/acs.chemmater.2c01976}
}

@article{5,
  author  = {Barai, Anup and Uddin, Khawaja and Dubarry, Matthieu and Somerville, Luke and McGordon, Andrew and Bloom, Ira},
  title   = {{A comparison of methodologies for the non-invasive characterisation of commercial Li-ion cells}},
  journal = {Progress in Energy and Combustion Science},
  volume  = {72},
  pages   = {1--31},
  year    = {2019},
  doi     = {10.1016/j.pecs.2019.01.001}
}

@article{6,
  author  = {Bloom, Ira and Jansen, Andrew N. and Abraham, Daniel P. and Knuth, Jason and Jones, Scott A. and Battaglia, Vincent S. and Henriksen, Gary L.},
  title   = {{Differential voltage analyses of high-power lithium-ion cells: I. Technique and application}},
  journal = {Journal of Power Sources},
  volume  = {139},
  number  = {1--2},
  pages   = {295--303},
  year    = {2005},
  doi     = {10.1016/j.jpowsour.2004.07.021}
}

@article{7,
  author  = {Ovejas, Victor J. and Cuadras, Angel},
  title   = {{Effects of cycling on lithium-ion battery hysteresis and overvoltage}},
  journal = {Scientific Reports},
  volume  = {9},
  pages   = {14875},
  year    = {2019},
  doi     = {10.1038/s41598-019-51474-5}
}

@article{8,
  author  = {Mercer, Matthew P. and Peng, Cheng and Soares, Catarina and Hoster, Harry E. and Kramer, Daniel},
  title   = {{Voltage hysteresis during lithiation/delithiation of graphite associated with meta-stable carbon stackings}},
  journal = {Journal of Materials Chemistry A},
  volume  = {9},
  pages   = {492--504},
  year    = {2021},
  doi     = {10.1039/D0TA10403E}
}

@article{9,
  author  = {Kargl, Philipp and Drews, Valentin and Daubinger, Philipp and others},
  title   = {{Investigation of voltage and expansion hysteresis of Si-alloy-C/NMC622 pouch cells using dilatometry}},
  journal = {Journal of Power Sources},
  volume  = {548},
  pages   = {232042},
  year    = {2022},
  doi     = {10.1016/j.jpowsour.2022.232042}
}

@article{10,
  author  = {Jahn, Lukas and M{\"o}{\ss}le, Patrick and R{\"o}der, Felix and Danzer, Michael A.},
  title   = {{A physically motivated voltage hysteresis model for lithium-ion batteries using a probability distributed equivalent circuit}},
  journal = {Communications Engineering},
  volume  = {3},
  number  = {1},
  pages   = {74},
  year    = {2024},
  doi     = {10.1038/s44172-024-00221-4}
}

@article{11,
  author  = {Weng, Anqi and Siegel, Jason B. and Stefanopoulou, Anna},
  title   = {{Differential voltage analysis for battery manufacturing process control}},
  journal = {Frontiers in Energy Research},
  volume  = {11},
  pages   = {1087269},
  year    = {2023},
  doi     = {10.3389/fenrg.2023.1087269}
}

@article{12,
  author  = {Edge, Joshua S. and O'Kane, Simon and Prosser, Rob and others},
  title   = {{Lithium-ion battery degradation: what you need to know}},
  journal = {Physical Chemistry Chemical Physics},
  volume  = {23},
  number  = {14},
  pages   = {8200--8221},
  year    = {2021},
  doi     = {10.1039/D1CP00359C}
}

@article{13,
  author  = {Oney, G. and Monaco, F. and Mitra, S. and Medjahed, A. and Burghammer, M. and Karpov, D. and Mirolo, M. and Drnec, J. and Jolivet, I. C. and Arnoux, Q. and Tardif, S. and Jacquet, Q. and Lyonnard, S.},
  title   = {{Dead, Slow and Overworked Graphite: Operando X-ray Microdiffraction Mapping of Aged Electrodes}},
  journal = {Advanced Energy Materials},
  volume  = {15},
  number  = {38},
  pages   = {e02032},
  year    = {2025},
  doi     = {10.1002/aenm.202502032}
}

@article{14,
  author  = {Yang, Jun and Li, Yufei and Mijailovic, Aleksandar and Wang, Gang and Xiong, Jie and Mathew, K. and Lu, Wei and Sheldon, Brian W. and Wu, Qingyun},
  title   = {{Gradient porosity electrodes for fast charging lithium-ion batteries}},
  journal = {Journal of Materials Chemistry A},
  volume  = {10},
  number  = {22},
  pages   = {12114--12124},
  year    = {2022},
  doi     = {10.1039/D2TA01707E}
}

@article{15,
  author  = {Schott, Thomas and Robert, Romain and Ulmann, Pierre A. and Lanz, Patrick and Z{\"u}rcher, Simon and Spahr, Michael E. and Nov{\'a}k, Petr and Trabesinger, Samuel},
  title   = {{Cycling Behavior of Silicon-Containing Graphite Electrodes, Part A: Effect of the Lithiation Protocol}},
  journal = {The Journal of Physical Chemistry C},
  volume  = {121},
  number  = {34},
  pages   = {18423--18429},
  year    = {2017},
  doi     = {10.1021/acs.jpcc.7b05919}
}

@article{16,
  author  = {Gao, Hong and Xiao, Liang and Pl{\"u}mel, Ingo and Xu, Guo-Liang and Ren, Yang and Zuo, Xiaodong and Liu, Yijin and Schulz, Christian and Wiggers, Hartmut and Amine, Khalil and Chen, Zhenguo},
  title   = {{Parasitic Reactions in Nanosized Silicon Anodes for Lithium-Ion Batteries}},
  journal = {Nano Letters},
  volume  = {17},
  number  = {3},
  pages   = {1512--1519},
  year    = {2017},
  doi     = {10.1021/acs.nanolett.6b04551}
}

\end{document}